\documentclass[amsmath,amssymb,nofootinbib,twocolumn]{revtex4}
\usepackage{graphicx}
\usepackage{pgfplots}
\pgfplotsset{compat=1.17}
\usepackage{dcolumn}
\usepackage{color}
\usepackage{scrextend}
\usepackage{subfig}
\usepackage{bm}
\usepackage{amsmath,amssymb,graphicx}
\usepackage{epsfig}
\usepackage{enumerate}
\usepackage{array}
\usepackage{multirow}
\usepackage{tikz}
\usepackage{ragged2e}
\usepackage{textgreek}
\usepackage[normalem]{ulem}

\begin{document}
\title
{Topological sensing of superfluid rotation\\ using non-Hermitian optical dimers}
\author{Aritra Ghosh\footnote{aritraghosh500@gmail.com}, Nilamoni Daloi, and M. Bhattacharya}
\affiliation{School of Physics and Astronomy, Rochester Institute of Technology, 84 Lomb Memorial Drive, Rochester, New York 14623, USA}
\vskip-2.8cm
\date{\today}
\vskip-0.9cm

\vspace{5mm}
\begin{abstract}
We theoretically investigate a non-Hermitian optical dimer whose parameters are renormalized by dispersive and dissipative backaction from the coupling of the passive cavity with a ring-trapped Bose-Einstein condensate. The passive cavity is driven by a two-tone control laser, where each tone is in a coherent superposition of Laguerre-Gaussian beams carrying orbital angular momenta $\pm \ell \hbar$. This imprints an optical lattice on the ring trap, leading to Bragg-diffracted sidemode excitations. Using an exact Schur-complement reduction of the full light-matter dynamics, we derive a frequency-dependent self-energy and identify a static regime in which the atomic response produces a complex shift of the passive optical mode. This renormalized dimer supports a tunable exceptional point, enabling spectroscopic signatures in the optical transmission due to a probe field, which can in turn be utilized for estimating the winding number of the persistent current. Exploiting the associated half-integer topological charge, we propose a digital exceptional-point-based sensing scheme based on eigenmode permutation, providing a noise-resilient method to sense superfluid rotation without relying on fragile eigenvalue splittings. Importantly, the sensing proposals are intrinsically nondestructive, preserving the coherence of the atomic superfluid.
\end{abstract}

\maketitle

\section{Introduction}
The study of non-Hermitian systems has emerged as a powerful framework for open systems in which gain and loss play an essential role \cite{Rotter_2009,ElGanainy_2018}. A striking feature of such systems is the occurrence of exceptional points, which are non-Hermitian degeneracies at which both eigenvalues and eigenvectors coalesce \cite{Berry_2004,Heiss_2012}. In atomic, molecular, and optical physics, exceptional points have not only been observed experimentally \cite{Kim_2016,Liang_2023}, but have also been utilized for sensors whose response to small perturbations is enhanced by the characteristic square-root splitting of the eigenvalues in their vicinity \cite{Wiersig_2014,Wiersig_2016,Chen_2017,Wiersig_2020}. A series of works, however, has clarified that the same mechanism also amplifies technical and quantum noise, severely limiting sensing advantage \cite{Wiersig_2020,Lau_2018,Langbein_2018}. These insights have motivated the search for exceptional-point-based sensing strategies that retain the topological robustness \cite{Heiss_2012} of non-Hermitian degeneracies while avoiding reliance on continuous eigenvalue splittings \cite{Doppler_2016}.

\vspace{2mm}

Non-Hermitian optical dimers are prototypical systems that may exhibit exceptional points as well as $\mathcal{PT}$-symmetry \cite{Ruter_2010}. These remarkable systems, consisting of two coherently-coupled cavity or waveguide modes, have found a variety of applications, including laser engineering \cite{Peng_2014}, optical isolation and nonreciprocal transport \cite{Chang_2014}, and sensing \cite{Wiersig_2014,Chen_2017}. Such diverse applications clearly illustrate how even the simplest two-mode optical structures can function as versatile building blocks for state-of-the-art platforms. Cavity platforms provide remarkable testbeds for studying light-matter interactions \cite{Kippenberg_2007,Aspelmeyer_2014,Weis_2010}, also opening up new directions of research involving ultracold atoms \cite{Brennecke_2008,Ritsch_2013}. A promising setup is provided by ring-trapped Bose-Einstein condensates (BECs) coupled to Fabry-P\'erot cavities \cite{Kumar_2021}. Such setups have been theoretically investigated in the context of the detection of solitons \cite{1_Pradhan_2024}, rotation sensing \cite{2_Pradhan_2024,Gupta_2024}, and Andreev-Bashkin effect \cite{Pradhan_2025}, among other applications. 

\vspace{2mm}

In this work, we shall exploit this versatile platform to theoretically investigate a non-Hermitian optical dimer that is renormalized by its coupling to a ring-trapped BEC. By considering two coupled cavities, one passive and one active, whose bare gain-loss balance would ordinarily give rise to familiar $\mathcal{PT}$-symmetric dimer physics, we will show how the inclusion of a ring-trapped BEC in the passive cavity leads to an effective non-Hermitian optical dimer. Compared to the bare optical dimer, the one that incorporates the BEC experiences dispersive and dissipative renormalization of the effective parameters due to cavity-assisted light-matter coupling in the passive cavity. Using a Schur-complement reduction, we will derive an exact frequency-dependent self-energy and identify a static regime in which the BEC backaction reduces to a complex, detuning-controlled shift of the passive mode. This shall allow us to obtain analytic conditions for the existence of exceptional points in the renormalized optical dimer where the dimer supermodes coalesce. The existence of exceptional points will then be utilized to put forward sensing proposals to determine superfluid rotation. 

\vspace{2mm}

Let us now present the organization of this paper. The details of the theoretical model will be discussed in Sec. (\ref{model_sec}) in which we shall also set up our notation and conventions. Then, in Sec. (\ref{NH_sec}), we will describe the effective non-Hermitian description which arises due to environmental loss and engineered gain, eventually leading to the identification of an exceptional point in the parameter space in Sec. (\ref{EP_sec}). This will allow us to present a proposal for estimating the winding number of the persistent current from the transmission spectrum. Moreover, exploiting the non-Hermitian topology of the exceptional point, in Sec. (\ref{topology_sec}), we shall propose a topological-sensing scheme for the winding number of the atomic persistent current. Finally, we shall conclude the paper in Sec. (\ref{conc_sec}). 

\section{Theoretical model}\label{model_sec}
\begin{figure}
\centering
\includegraphics[width=0.9\linewidth]{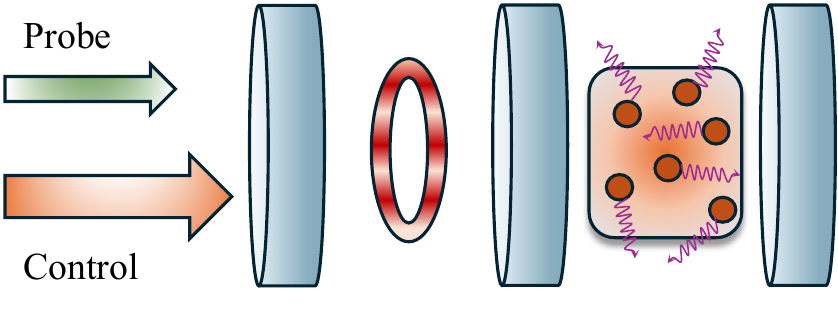}
\caption{\justifying{Schematic setup showing the two optical cavities coupled evanescently. The cavity on the left side is the passive cavity with loss rate $\gamma_0$ that contains the ring-trapped BEC and is controlled by a two-tone control laser where each tone is in a coherent superposition of Laguerre-Gaussian modes carrying OAM $\pm \ell \hbar$. The active cavity on the right admits a net gain rate $\Gamma = g_0 - \gamma'$, where $\gamma'$ is the intrinsic loss rate of this cavity and $g_0$ is the gain rate due to the active medium. A probe field is later included for spectroscopic readout.}}
\label{schematic}
\end{figure}

We shall consider two Fabry-P\'erot cavities, one of which has a net optical damping $\gamma_0$ while the other admits a net optical gain $\Gamma$. Additionally, we will put in the passive cavity, a BEC of $N$ identical $^{23}{\rm Na}$ atoms of mass $m$, confined in an annular ring trap \cite{Morizot_2006,Wright_2013} of radius $R_0$ and potential $V(\rho) = \frac{1}{2}m\omega_\rho^2 (\rho-R_0)^2$, as illustrated in Fig. (\ref{schematic}). The atoms undergo quantized rotational motion around the cavity axis, characterized by a winding number $L_p \in \mathbb{Z}$~\cite{Wright_2013} and rotational energy $\hbar^2 L_p^2/(2mR_0^2)$~\cite{Kumar_2021}. Focusing now on the passive cavity, it is driven by two coherent
control tones at frequencies $\omega_{L1}$ and $\omega_{L2}$ with complex drive strengths $\varepsilon_1$ and $\varepsilon_2$. Both the tones populate the same intracavity optical mode described by the bosonic operators $(a,a^\dagger)$, and each tone is prepared in a coherent superposition of Laguerre-Gaussian modes \cite{Molina-Terriza_2001,Yao_2011,Fickler_2012} carrying orbital angular momenta (OAM) $\pm\ell\hbar$, thereby generating a circular optical lattice overlapping with the ring-shaped BEC. It may be emphasized that of interest to us is a single longitudinal cavity resonance, while the other longitudinal resonances are separated by the free spectral range and are therefore detuned.

\vspace{2mm}

In the rotating frame of the second control tone, the driven (passive) cavity Hamiltonian is
\begin{equation}
\frac{H_{\rm pc}}{\hbar}= -\Delta_2a^\dagger a
+i(\varepsilon_2 a^\dagger-\varepsilon_2^* a)
+i(\varepsilon_1 e^{i\delta_0 t} a^\dagger-\varepsilon_1^* e^{-i\delta_0 t} a),
\end{equation}
where the subscript `pc' stands for passive cavity, $\Delta_2=\omega_{L2}-\omega_0$, and $\delta_0=\omega_{L2}-\omega_{L1}$. The atomic Hamiltonian on the ring is
\begin{eqnarray}
H_{\rm ring}&=&\int_0^{2\pi} d\phi \Psi^\dagger(\phi) \mathcal{H} \Psi(\phi)\\
&&+ \frac{g}{2} \int_{0}^{2\pi} d\phi \Psi^{\dagger}(\phi)\Psi^{\dagger}(\phi)\Psi(\phi)\Psi(\phi), \nonumber\\
\mathcal{H} &=& -\frac{\hbar^2}{2mR_0^2}\frac{\partial^2}{\partial\phi^2}+\hbar U_0 \cos^2(\ell\phi)a^\dagger a, \nonumber
\end{eqnarray}
where $\Psi(\phi)$ is the atomic field operator that satisfies $[\Psi(\phi),\Psi^{\dagger}(\phi')]=\delta(\phi-\phi')$, $g=2\hbar\omega_{\rho}a_{\rm Na}/R_0$ is the effective interatomic-interaction strength with $a_{\rm Na}$ being sodium's $s$-wave scattering length, and $U_0$ is the single-photon dispersive light shift. The optical lattice induces Bragg scattering between rotational states whose winding numbers differ by $2\ell$. The optical lattice will be taken to be weak \cite{Kumar_2021}, so that by retaining only the lowest-order diffraction effects, the atomic field can be expanded as
\begin{equation}
\Psi(\phi) = \frac{1}{\sqrt{2\pi}}
\Big[e^{iL_p\phi} c_p + e^{i(L_p+2\ell)\phi} c_+ + e^{i(L_p-2\ell)\phi} c_-\Big],
\end{equation}
with bosonic operators $c_{p,\pm}$ satisfying
$c_p^\dagger c_p+c_+^\dagger c_+ + c_-^\dagger c_-=N$.
Since the persistent-current mode $c_p$ is macroscopically occupied, we shall treat it classically ($c_p^\dagger c_p\simeq N$) and define the sidemode operators
\begin{equation}
c=\frac{c_p^\dagger c_+}{\sqrt{N}}, \quad \quad d=\frac{c_p^\dagger c_-}{\sqrt{N}},
\end{equation}
which satisfy $[c,c^\dagger]=[d,d^\dagger]=1$ for large $N$.
The resulting Hamiltonian describing the optical field and two atomic sidemodes is
\begin{eqnarray}
\frac{H_{\rm pc+ring}}{\hbar}
&=& -\tilde\Delta_2a^\dagger a +\omega_cc^\dagger c +\omega_dd^\dagger d +G(X_c+X_d)a^\dagger a \nonumber\\
&&+i(\varepsilon_2 a^\dagger-\varepsilon_2^* a)
+i(\varepsilon_1 e^{i\delta_0 t} a^\dagger-\varepsilon_1^* e^{-i\delta_0 t} a) \nonumber\\
&&+ 4\tilde{g}N(c^\dagger c + d^\dagger d)
+ 2\tilde{g}N(c d + c^\dagger d^\dagger),
\end{eqnarray}
where $X_{c(d)}=(c_{(d)}+c_{(d)}^\dagger)/\sqrt{2}$,
$\omega_{c(d)}=\hbar[L_p\pm2\ell]^2/(2mR_0^2)$,
$G=U_0\sqrt{N/8}$, $\tilde\Delta_2=\Delta_2-U_0N/2$, and $\tilde{g}=g/(4\pi\hbar)$ denotes the strength of interatomic interactions. The truncation to the $\pm 2\ell$ Bragg modes follows from the angular-momentum selection rule imposed by the lattice potential $\cos^2(\ell\phi)$, with nonzero Fourier components at $0$ and $\pm 2\ell$. In the weak-lattice regime of interest here, the leading-order diffraction couples the macroscopically-occupied persistent-current mode $L_p$ to $L_p\pm 2\ell$, while higher-order diffraction processes are suppressed by higher powers of the lattice depth. The interatomic-interaction-induced corrections are negligible in the parameter regime that we shall work with ($\omega_{c,d} \gg \tilde{g}N)$ and therefore can be dropped. The above-mentioned form of the Hamiltonian incorporates optomechanical-type coupling between the atomic (mechanical) sidemodes and the intracavity mode. 

\vspace{2mm}

Let us linearize the Hamiltonian by writing
\begin{equation}
a(t)=\bar{a}(t)+\tilde{a},
\quad \quad
c=\alpha_c+\tilde{c},
\quad \quad
d=\alpha_d+\tilde{d},
\end{equation}
where the intracavity mean field contains both control tones as
$\bar{a}(t)=\alpha_2+\alpha_1 e^{i\delta_0 t}$. Keeping fluctuation terms up to the second order and using the classical
equations of motion to eliminate linear terms yields
\begin{eqnarray}
\frac{H^{\rm lin}_{\rm pc+ring}}{\hbar} &=&
-\bar{\Delta}\tilde{a}^\dagger \tilde{a}
+\omega_c\tilde{c}^\dagger \tilde{c}
+\omega_d\tilde{d}^\dagger \tilde{d} \\
&&+G(\tilde{X}_c+\tilde{X}_d)\big[\bar{a}^*(t)\tilde{a}+\bar{a}(t)\tilde{a}^\dagger\big], \nonumber
\end{eqnarray}
with $\bar{\Delta}\approx\tilde{\Delta}_2$ as the light-matter-coupling-induced shift is negligible. The linearized optomechanical interaction inherits an explicit time dependence from the two-tone intracavity field $\bar{a}(t)=\alpha_2+\alpha_1 e^{i\delta_0 t}$. Moving to the interaction picture with respect to the free Hamiltonian $H_0 = -\hbar\bar{\Delta} \tilde{a}^\dagger \tilde{a}
+ \hbar \omega_c \tilde{c}^\dagger \tilde{c}
+\hbar \omega_d \tilde{d}^\dagger \tilde{d}$ and choosing the detuning and tone separation such that $-\bar{\Delta}\simeq\omega_d$ and $-\bar{\Delta}+\delta_0\simeq\omega_c$ (i.e., $\delta_0\simeq\omega_c-\omega_d$), each control tone becomes resonant with the red sideband of a distinct atomic sidemode: the tone at $\omega_{L2}$ couples the cavity mode to sidemode $d$, while the tone at $\omega_{L1}$ couples it to sidemode $c$. In the resolved-sideband regime $\omega_{c,d}\gg\gamma_0, |G\alpha_{1,2}|$, the two-mode-squeezing terms oscillate rapidly and average out, and the rotating-wave approximation yields a time-independent effective Hamiltonian containing only beam-splitter interactions with equal coupling strengths $\tilde{G}$ set by appropriately choosing the control-tone amplitudes. One thus arrives at the following effective Hamiltonian addressing the passive cavity including the ring-trapped BEC:
\begin{eqnarray}
\frac{H^{\rm eff}_{\rm pc+ring}}{\hbar}
&=& -\bar{\Delta} a^\dagger a +\omega_cc^\dagger c +\omega_dd^\dagger d \label{Heffpcring} \\
&&+\tilde{G}(a^\dagger c+a c^\dagger)
+\tilde{G}(a^\dagger d+a d^\dagger), \nonumber
\end{eqnarray}
where we have relabeled $(\tilde{a},\tilde{c},\tilde{d})\to(a,c,d)$ for simplicity. Let us re-emphasize the validity of this reduction based on the standard approach of treating Bose-Einstein condensates in optical potentials in cavities \cite{Brennecke_2008,Ritsch_2013,Kumar_2021}. First, we have worked in the weak-lattice regime where the dominant angular-momentum transfer induced by the lattice $\cos^2(\ell\phi)$ is $\pm 2\ell$, so that higher-order diffraction processes are suppressed by higher powers of the lattice depth. Second, we have assumed the resolved-sideband hierarchy $\omega_{c,d} \gg \gamma_0,\ |\tilde{G}|$, so that the counter-rotating terms oscillate rapidly and average out. Third, by selecting the tone separation $\delta_0\simeq \omega_c-\omega_d$ and detunings $-\bar\Delta\simeq \omega_d$, $-\bar\Delta+\delta_0\simeq \omega_c$, one can ensure that each tone addresses the red sideband of a distinct sidemode, while off-resonant cross-couplings are suppressed provided $|\omega_c-\omega_d| \gg \gamma_0,\ |\tilde{G}|$. Finally, we have restricted to a weak-interaction regime $\omega_{c,d}\gg \tilde{g} N$, so that interaction-induced collective-mode renormalizations do not modify the relevant sidemode frequencies on the scales of interest. Our parameter values are $\gamma_{0}=2\pi~{\rm kHz}$, $\omega_c = 40.04\gamma_{0}$, $\omega_d = 19.83\gamma_{0}$, $\tilde{G} = 2\gamma_{0}$, and $4\tilde{g}N \approx 0.05\gamma_0$, which respect all the inequalities required for these approximations to consistently hold.

\vspace{2mm}

Including now the active cavity which couples evanescently with the passive cavity, the full linearized Hamiltonian takes the form 
\begin{eqnarray}
\frac{H}{\hbar}
&=& -\bar{\Delta} (a^\dagger a + b^\dagger b) + \omega_c c^\dagger c + \omega_d d^\dagger d  \nonumber \\
&& ~ + \tilde{G}(a^\dagger c + a c^\dagger) + \tilde{G}(a^\dagger d + a d^\dagger) \nonumber\\
&&~+ J(a^\dagger b + ab^\dagger), 
\label{H_full}
\end{eqnarray}
where $J$ is the effective evanescent-coupling constant, taken real by phase choice, and $(b,b^\dagger)$ are the operators for the active-cavity fluctuations. Note that one can choose the resonance frequency of the active cavity such that in a frame rotating with respect to the passive cavity's control laser (second tone), $-\bar{\Delta} b^\dagger b$ represents the active-cavity Hamiltonian, supplemented by the tunneling interaction between the two cavities, i.e., the effective detuning for mode $b$ is chosen to match that of $a$ via cavity design. 

\section{Non-Hermitian description}\label{NH_sec}
The Heisenberg equations from the Hamiltonian (\ref{H_full}) for the four modes read
\begin{eqnarray}
\dot{a} &=& \bigg(i\bar{\Delta} - \frac{\gamma_0}{2}\bigg)a - i J b- i\tilde{G}(c+d), \\
\dot{b} &=& \bigg(i\bar{\Delta} + \frac{\Gamma}{2}\bigg) b - i J a, \\
\dot{c}  &=& \bigg(-i\omega_c - \frac{\gamma_m}{2}\bigg) c - i\tilde{G} a, \\
\dot{d}   &=& \bigg(-i\omega_d - \frac{\gamma_m}{2}\bigg)d - i\tilde{G} a,
\end{eqnarray} up to noises that have not been made explicit above and the standard damping rates have been included. Note that the mode $b$ is antidamped with rate $\Gamma$. These equations can be cast in matrix form in the manner
\begin{equation}
\dot{A} = i \Lambda A + A_{\rm in},
\end{equation} where $A = (a~b~c~d)^T$ and the deterministic part of the time evolution is governed by the non-Hermitian matrix 
\begin{equation}
\Lambda =
\begin{pmatrix}
\bar{\Delta} + i\dfrac{\gamma_0}{2} & -J & -\tilde{G} & -\tilde{G} \\[6pt]
-J & \bar{\Delta} - i\dfrac{\Gamma}{2} & 0 & 0 \\[6pt]
-\tilde{G} & 0 & -\omega_c + i\dfrac{\gamma_m}{2} & 0 \\[6pt]
-\tilde{G} & 0 & 0 & -\omega_d + i\dfrac{\gamma_m}{2}
\end{pmatrix}. 
\label{Lambdadef}
\end{equation}
Since there is no obvious balance of gain and loss, the quantum dynamics is generally not $\mathcal{PT}$-symmetric. 

\subsection{Reduction to the optical subspace}
The $4 \times 4$ problem identified above can be simplified to a $2 \times 2$ problem by projecting the atomic effects onto the optical subspace spanned by the operators $a$ and $b$. A direct calculation invoking the Schur-complement reduction (see Appendix (\ref{appA})) allows one to define an effective optical matrix that goes as
\begin{equation}
M_{\rm eff}(\lambda)=
\begin{pmatrix}
\bar{\Delta}+i\dfrac{\gamma_0}{2}+\Sigma(\lambda) & -J\\[6pt]
-J & \bar{\Delta}-i\dfrac{\Gamma}{2}
\end{pmatrix},
\label{Mefflambda}
\end{equation}
where $\lambda$ satisfies the characteristic equation of the matrix (\ref{Lambdadef}) and one has a complex self-energy
\begin{equation}
\Sigma(\lambda) = \dfrac{\tilde{G}^2}{\lambda+\omega_c-i\dfrac{\gamma_m}{2}} + \dfrac{\tilde{G}^2}{\lambda+\omega_d-i\dfrac{\gamma_m}{2}},
\label{selfenergy}
\end{equation} interpreted as the atom-induced shift to the optical modes. The real part of $\Sigma(\lambda)$ gives a Lamb shift of the passive mode $a$, while its imaginary part modifies the effective loss or gain balance between $a$ and $b$. The latter implies that if we started with a bare optical dimer with balanced gain and loss, i.e., $\Gamma = \gamma_0$, the atomic backaction makes the optical dimer unbalanced. The exact form of $\Sigma(\lambda)$ contains poles near the atomic-sidemode frequencies $\lambda \simeq -\omega_{c,d}$. To work with a closed $2\times 2$ optical matrix, it is convenient to replace $\Sigma(\lambda)$ by its static value $\Sigma(\bar{\Delta})$ evaluated at the control detuning. This static approximation is justified whenever the self-energy varies slowly across the optical-eigenvalue window.  Physically, this means the optical modes must lie several linewidths away from the atomic sidemodes so that the atomic susceptibility is not sampled over the optical bandwidth.  Importantly, while the poles at $-\omega_{c,d}$ are sharply peaked on the tiny scale $\gamma_{m}/2\sim10^{-5}\gamma_{0}$ (in our choice of parameters), violation of the static approximation occurs only if the optical eigenvalues are tuned into the vicinity of these poles, in which case the full $\lambda$-dependence of $\Sigma(\lambda)$ must be retained. We shall restrict our attention to this static regime (see Appendix (\ref{appB}) for more details on its validity) which leads to the static effective matrix
\begin{equation}
M_{\rm eff}(\bar{\Delta})\approx
\begin{pmatrix}
\bar{\Delta}+i\dfrac{\gamma_0}{2}+\Sigma(\bar{\Delta}) & -J\\[6pt]
-J & \bar{\Delta}-i\dfrac{\Gamma}{2}
\end{pmatrix}.
\label{Meff_static}
\end{equation}
This form captures the leading-order influence of the atomic modes as a complex renormalization. It is noteworthy that one must choose $\bar{\Delta}$ such that the optical eigenvalues remain spectrally separated from the atomic poles at $\lambda\simeq-\omega_{c,d}$, ensuring that the atomic backaction enters only through the off-resonant self-energy $\Sigma(\bar{\Delta})$ within the static approximation.
\begin{figure}
\centering
\includegraphics[width=0.9\linewidth]{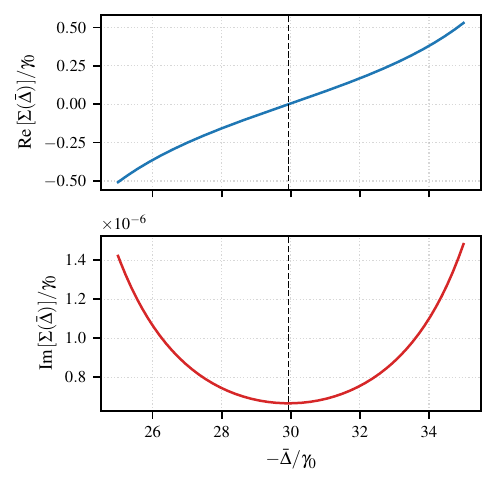}
\caption{\justifying{Real and imaginary parts of $\Sigma(\bar{\Delta})$. The parameters are $\omega_c = 40.04\gamma_{0}$, 
$\omega_d = 19.83\gamma_{0}$, 
$\tilde{G} = 2\gamma_{0}$, 
and $\gamma_m = 1.7\times 10^{-5}\gamma_{0}$, with $\gamma_{0}=2\pi~{\rm kHz}$. The sidemode frequencies are obtained by putting $m = 23$ amu, $R_0 = 10~\mu$m, $L_p = 115$, and $\ell = 10$ in $\omega_{c(d)} = \frac{\hbar[L_p + (-) 2\ell]^2}{2mR_0^2}$. The dashed vertical line corresponds to $\bar{\Delta}_{0}=-(\omega_c+\omega_d)/2\simeq-29.94\gamma_{0}$, where the real part changes sign.}}
\label{sigma}
\end{figure}
The real and imaginary parts of $\Sigma(\bar{\Delta})$ are shown in Fig. (\ref{sigma}), and admit the analytical expressions
\begin{eqnarray}
{\rm Re}[\Sigma(\bar{\Delta})] &=& \tilde{G}^2 \bigg[\frac{\bar{\Delta} + \omega_c}{\chi_c} + \frac{\bar{\Delta} + \omega_d}{\chi_d} \bigg], \label{ReSigma} \\
{\rm Im}[\Sigma(\bar{\Delta})] &=& \tilde{G}^2 \bigg(\frac{\gamma_m}{2}\bigg) \bigg[\frac{1}{\chi_c} + \frac{1}{\chi_d} \bigg], \label{ImSigma}
\end{eqnarray} where $\chi_{c,d} = (\bar{\Delta} + \omega_{c,d})^2 + (\gamma_m/2)^2$. Solving the characteristic equation of the matrix (\ref{Meff_static}) leads to the eigenvalues
\begin{eqnarray}
\lambda_\pm &=& \bar{\Delta} + \frac{{\rm Re}[\Sigma(\bar{\Delta})]}{2}
+ i\left(\frac{\gamma_0-\Gamma}{4} + \frac{{\rm Im}[\Sigma(\bar{\Delta})]}{2}\right) \nonumber \\
&&~~\pm \frac{1}{4}
\sqrt{16J^2 + \big(i(\gamma_0+\Gamma) + 2\Sigma(\bar{\Delta})\big)^2}. 
\label{eigenvalues_general}
\end{eqnarray} 
The eigenvalues are generally complex-valued even if the discriminant is real and positive. These eigenvalues correspond to the dimer supermodes which exist in a superposition of the $a$ and $b$ optical modes. It is noteworthy that the reality of the eigenvalues can be obtained if two conditions are met simultaneously: (i) the discriminant under the square root in the expression (\ref{eigenvalues_general}) is real and positive-semidefinite, and (ii) the renormalized gain-loss balance
\begin{equation}
\Gamma = \gamma_0 + 2 {\rm Im}[\Sigma(\bar{\Delta})],
\label{renormalized_gainlossbalance}
\end{equation} is imposed. Of course, in the case of the bare optical dimer, it reduces to the familiar $\Gamma = \gamma_0$. 

\subsection{Observable signatures in optical transmission}
\begin{figure}
\centering
\includegraphics[width=0.9\linewidth]{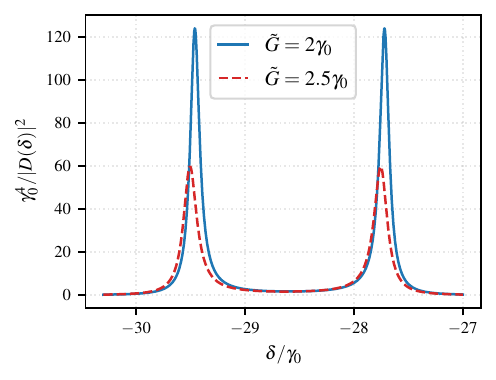}
\caption{\justifying{Transmission proxy $(\gamma_{0}^{2}/|D(\delta)|)^{2}$ as a function of the probe detuning $\delta/\gamma_{0}$, calculated from the effective non-Hermitian optical dimer including atomic backaction, for two different values of $\tilde{G}$. The remaining parameters are fixed to $\bar{\Delta}=-27\gamma_{0}$, $J=\gamma_{0}$, $\Gamma=\gamma_{0}$, $\gamma_{m}=1.7\times 10^{-5}\gamma_{0}$, and atomic-sidemode frequencies $\omega_{c}=40.04\gamma_{0}$ and $\omega_{d}=19.83\gamma_{0}$.}}
\label{D_delta}
\end{figure}

The non-Hermitian nature of the optical supermodes can be probed directly via pump-probe spectroscopy \cite{Kippenberg_2007,Weis_2010}. In the static regime discussed above, the optical fields $(a,b)$ evolve under the effective matrix (\ref{Meff_static}), where all atomic-backaction effects enter through the complex self-energy $\Sigma(\bar{\Delta})$. Let us say a weak probe field at frequency $\omega_p$ is injected into cavity~$a$, corresponding to a detuning $\delta=\omega_{L2}-\omega_p$ from the second control tone in the rotating frame of the latter. In the frequency space, the steady-state fields satisfy
\begin{equation}
\big(M_{\rm eff}-\delta I\big)
\begin{pmatrix}
a(\delta)\\[2pt] b(\delta)
\end{pmatrix}
=
\begin{pmatrix}
\eta\\[2pt] 0
\end{pmatrix},
\label{probe_eq_expt}
\end{equation}
with probe amplitude $\eta$ applied to cavity $a$. Considering $b$, one finds the exact expression 
\begin{equation}
b(\delta) = \eta\frac{J}{D(\delta)}, \quad \quad D(\delta) = \det\big(M_{\rm eff}-\delta I\big).
\label{probe_sol_expt}
\end{equation}
The same susceptibility denominator $D(\delta)$ is obtained even if the input noises are included within the quantum Langevin framework (see Appendix (\ref{appC})). The transmitted field from cavity~$b$ follows from standard input-output relations $b_{\rm out}(\delta) \propto b(\delta)$, leading to the following transmission intensity at the probe frequency:
\begin{equation}
T_b(\delta) \propto |b(\delta)|^2 = |\eta|^2 \frac{|J|^2}{|D(\delta)|^2}.
\label{T_b}
\end{equation}
The experimentally-measured spectrum is therefore governed entirely by the inverse modulus of $D(\delta)$ which contains information about the supermode eigenvalues (\ref{eigenvalues_general}). Thus the quantity $\big(\gamma_{0}^{2}/|D(\delta)|\big)^{2}$ can be used as a dimensionless proxy for the probe transmission since the measured transmission from cavity $b$ is proportional to $|D(\delta)|^{-2}$ up to an overall coupling-dependent factor. The quantity $|D(\delta)|^{-2}$ has been depicted in Fig. (\ref{D_delta}) in dimensionless form exhibiting the transmission peaks.  
Since $D(\delta)=0$ is equivalent to 
$\delta=\lambda_\pm$, writing these eigenvalues as $\lambda_\pm=\Omega_\pm + i\frac{\kappa_\pm}{2}$ (with $\kappa_\pm = 2{\rm Im}[\lambda_\pm]$), one immediately sees that (a) the resonance peak positions occur near 
$\delta={\rm Re}[\lambda_\pm]$, (b) the linewidths of the peaks are governed by the imaginary parts $|\kappa_\pm|$, and (c) the splitting of the resonances is given by $\Delta\Omega = {\rm Re}(\lambda_+ - \lambda_-)$, directly resolvable in the transmission spectrum. 

\vspace{2mm}

Because atomic backaction directly impacts mode $a$, which is in turn coupled to $b$, the resulting transmission spectrum through cavity~$b$ carries experimentally-accessible signatures of the atom-induced modification of the optical dimer. Unless the control detuning is taken so that ${\rm Re}[\Sigma(\bar{\Delta})] \simeq 0$, the imaginary part of the self-energy is much smaller than its real part, so the dominant effect is dispersive. The real part of $\Sigma(\bar{\Delta})$ manifests as a clear shift or deformation of the frequency separation between the two peaks, providing a direct spectroscopic probe of dispersive atomic backaction. Thus the transmission coefficient (\ref{T_b}) can establish a direct and quantitative link between the complex-valued eigenstructure of $M_{\rm eff}(\bar{\Delta})$ and the experimentally-measured transmission from cavity~$b$. 

\section{Exceptional points}\label{EP_sec}
\begin{figure}
\centering
\includegraphics[width=0.8\linewidth]{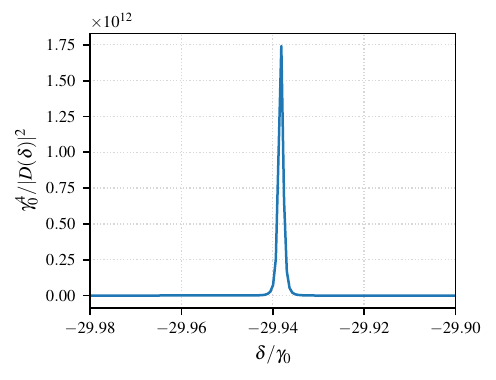}
\caption{\justifying{Transmission proxy $(\gamma_{0}^{2}/|D(\delta)|)^{2}$ as a function of the probe detuning $\delta/\gamma_{0}$ at the exceptional point. The parameters are $\tilde{G}=3\gamma_{0}$, $\Gamma=\gamma_{0}$, $\gamma_{m}=1.7\times 10^{-5}\gamma_{0}$, $\omega_{c}=40.04\gamma_{0}$, $\omega_{d}=19.83\gamma_{0}$, and $J=J_{\rm EP}$. Both the transmission peaks have coalesced into a single enhanced peak at $\delta \simeq -29.94\gamma_0$.}}
\label{D_delta_EP}
\end{figure}

\begin{figure}
\centering
\includegraphics[width=1\linewidth]{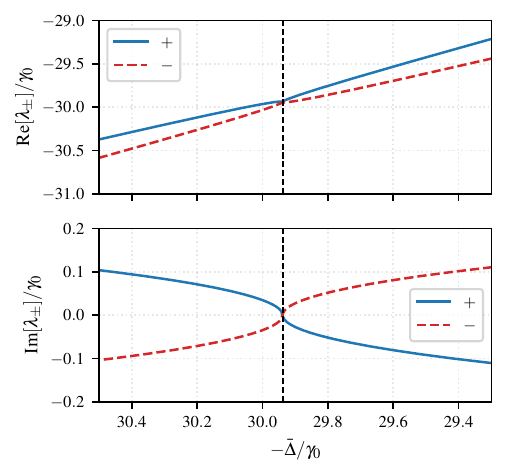}
\caption{\justifying{Real and imaginary parts of the eigenvalues $\lambda_\pm$ of $M_{\rm eff}(\bar{\Delta})$. The parameters are $\tilde{G}=2\gamma_{0}$, $\Gamma=\gamma_{0}$, $\gamma_{m}=1.7\times 10^{-5}\gamma_{0}$, $\omega_{c}=40.04\gamma_{0}$, $\omega_{d}=19.83\gamma_{0}$, and $J=J_{\rm EP}$ as given by expression (\ref{JEP_condition_physical}). The exceptional point is seen to occur at $\bar{\Delta} = - \frac{\omega_c + \omega_d}{2} \simeq -29.94\gamma_0$ (black-dashed line).}}
\label{eig_EP}
\end{figure}

Let us now explore the exceptional points. At an exceptional point, the complex-valued discriminant inside the square root of the eigenvalues (\ref{eigenvalues_general}) should vanish, requiring
\begin{equation}
\big[ 16J^2 + \big(i(\gamma_0+\Gamma) + 2\Sigma(\bar{\Delta})\big)^2\big]_{\rm EP} = 0. 
\label{EP_condition}
\end{equation}
Some algebra (see Appendix (\ref{appD}) for details) reveals that a nontrivial exceptional point which is consistent with the physical parameters can be obtained by choosing the control detuning to $\bar{\Delta} = \bar{\Delta}_0 = - \frac{\omega_c + \omega_d}{2}$, for which the real part of the complex self-energy vanishes. If the detuning is set to this value, an exceptional point is obtained for 
\begin{equation}
J_{\rm EP} \approx \frac{1}{4} \left(\gamma_0 + \Gamma + \frac{8 \tilde{G}^2 \gamma_m}{(\omega_c - \omega_d)^2}\right).
\label{JEP_condition_physical}
\end{equation}
In obtaining the above expression, we have used the fact that $|\omega_c - \omega_d| \gg \gamma_m$. The eigenvalues coalesce to a complex number which becomes real only if the renormalized gain-loss balance (\ref{renormalized_gainlossbalance}) is enforced. In Fig. (\ref{D_delta_EP}), we have demonstrated the behavior of the transmission proxy $(\gamma_{0}^{2}/|D(\delta)|)^{2}$ at the exceptional point which shows the merging of the peaks, as may be observed in an experiment.
Fig. (\ref{eig_EP}) depicts the eigenvalues $\lambda_\pm$, showing coalescence at the exceptional point. The parameters are set such that the condition (\ref{renormalized_gainlossbalance}) is not met, thereby leading to nontrivial imaginary parts on either side of the exceptional point. 

\vspace{2mm}

It is worth noting that the Schur-complement reduction is an exact algebraic elimination of the atomic subspace and gives rise to an exact energy-dependent optical problem. The existence of an exceptional point is thus determined by the defectiveness of the corresponding non-Hermitian linear operator, while the reduced formulation simply provides a convenient representation of this condition. The only approximation that we have invoked in locating the exceptional point analytically is the static replacement $\Sigma(\lambda)\to\Sigma(\bar{\Delta})$ in the regime quantified by the condition (\ref{static_condition}). In this controlled regime, the frequency dependence of $\Sigma(\lambda)$ may produce only a small perturbative shift in the exceptional-point location and does not introduce spurious degeneracies within the controlled off-resonant regime (see Appendix (\ref{appB})).

\subsection{Estimating $L_p$ from exceptional-point location}
Let us now put forward a simple proposal for sensing superfluid rotation, i.e., the winding number $L_p$, based on tracking the exceptional-point location. The idea rests on the fact that the control detuning's value that leads to the exceptional point has a strong dependence on the sidemode frequencies $\omega_{c,d}$, which in turn depend on $L_p$. Since for our typical parameters, if one begins with a gain-loss balanced optical dimer, $J_{\rm EP} \approx \gamma_0/2$, one can fabricate a two-cavity system with predefined $J = J_{\rm EP}$. For arbitrary values of the control detuning, the transmission spectrum shows two peaks. Thus by carefully varying the detuning, the peaks can be observed to coalesce at $\bar\Delta_0 = - \frac{\omega_c + \omega_d}{2}$, and from the value of this detuning $\bar\Delta = \bar\Delta_0$, one can determine 
\begin{equation}
L_p^2 = -\frac{2mR_0^2\bar\Delta_0}{\hbar} - 4\ell^2.
\end{equation} Since $m$ and $R_0$ are fixed numbers, while $\ell$ for the source is known, one can determine or `sense' $|L_p|$. 

\vspace{2mm}

The precision of this exceptional-point-based estimation is ultimately limited by the spectral linewidth with which the exceptional-point detuning $\bar{\Delta}_0$ can be realistically identified. Since $\bar{\Delta}_0$ depends on the winding number as
\begin{equation}
\bar{\Delta}_0=-\frac{\hbar}{2mR_0^2}\left(L_p^2+4\ell^2\right),
\end{equation}
an uncertainty $|\delta\bar{\Delta}_0|$ translates into an uncertainty in the inferred winding number in the manner $|\delta L_p| \simeq \frac{mR_0^2}{\hbar |L_p|}|\delta\bar{\Delta}_0|$. In a linewidth-limited measurement, the smallest-resolvable detuning shift is set by the effective linewidth $\kappa_{\rm EP}$ of the optical supermodes near the exceptional point, so that $|\delta\bar{\Delta}_0| \sim \kappa_{\rm EP}/2$. This immediately gives
\begin{equation}
|\delta L_p| \sim \frac{mR_0^2}{\hbar |L_p|}\frac{\kappa_{\rm EP}}{2}.
\end{equation}
For typical parameters $m=23$~amu, $R_0=10~\mu{\rm m}$, $L_p=115$, and $\kappa_{\rm EP} \sim \gamma_0 = 2\pi$ kHz, one finds $|\delta L_p| \sim \mathcal{O}(1)$ for the above-mentioned parameters with the precision improving for larger $|L_p|$. In other words, the exceptional-point-based estimation proposed above can perform particularly well in the large-$L_p$ regime. 

\vspace{2mm}

Unlike conventional exceptional-point sensors that infer a perturbation from the square-root splitting of eigenvalues and suffer from enhanced noise, the present estimation scheme relies on locating the control detuning at which the optical supermodes coalesce. As the estimator is based on locating the position of a spectral feature rather than resolving a small exceptional-point-induced eigenvalue splitting, its precision is governed primarily by the measurable linewidth of the transmission resonance and by technical noise sources such as frequency drift, gain noise, and parameter fluctuations. Consequently, the achievable sensitivity is set by ordinary spectral resolution rather than by the divergent susceptibility associated with non-Hermitian degeneracies.

\section{Topological sensing}\label{topology_sec}
\subsection{Topological charge}
Indicating the complex discriminant by $\mathfrak{D}(\bar{\Delta},J) = 16J^2 + \big[i(\gamma_0+\Gamma)+2\Sigma(\bar{\Delta})\big]^2$, an exceptional point occurs when $\mathfrak{D}(\bar{\Delta},J) = 0$. This determines the location of the exceptional point in the two-dimensional space of control parameters $\mathbf{R}=(\bar{\Delta},J)$. For the parameters $(\omega_{c,d},\tilde{G},\gamma_m)$ and the optical rates $(\gamma_0,\Gamma)$, the exceptional point is located within the static approximation at
\begin{eqnarray}
&&\Big(\bar{\Delta}_0(\omega_c), J_{\rm EP}(\omega_c)\Big)
= \label{REP_omegaC}\\
&&~~~~~~~\left(-\frac{\omega_c+\omega_d}{2}, \frac{1}{4}
\left(\gamma_0+\Gamma+\frac{8\tilde{G}^2\gamma_m}{(\omega_c-\omega_d)^2}\right)
\right), \nonumber
\end{eqnarray}
where we have used ${\rm Im}[\Sigma(\bar{\Delta}_0)] \simeq 4\tilde{G}^2\gamma_m/(\omega_c-\omega_d)^2$ in the regime $|\omega_c-\omega_d|\gg\gamma_m$. Let us denote $\mathbf{R}_{\rm EP} = (\bar{\Delta}_0(\omega_c), J_{\rm EP}(\omega_c))$ which is an isolated exceptional point in the two-dimensional parameter space $\mathbf{R}=(\bar{\Delta},J)$. In this case, the exceptional point behaves as a topological defect with a half-integer charge \cite{Heiss_2012,Berry_2004} (see Appendix (\ref{appE})):
\begin{equation}
q_{\rm EP} = \pm \frac{1}{2},
\label{halfint_charge}
\end{equation}
with the sign determined by the orientation of the loop encircling $\mathbf{R}_{\rm EP}$ once. This remarkable topological feature shall allow us to propose a topological scheme for sensing.

\subsection{Topologically-robust sensing}
We will now show how to implement topological sensing using the exceptional point. The key idea is to use the topological permutation of eigenmodes (see Appendix (\ref{appE})) upon encircling the exceptional point in a suitable two-dimensional control-parameter space as a robust binary observable instead of relying on continuous readout of eigenvalue splittings, naturally mitigating the well-known noise fragility of continuous exceptional-point-based sensing \cite{Wiersig_2020,Lau_2018,Langbein_2018}. The dependence of the exceptional-point location $\mathbf{R}_{\rm EP}(\omega_c)$ on the atomic-sidemode frequency $\omega_c$ naturally suggests the use of the exceptional point as a sensor for $\omega_c$. 

\vspace{2mm}

To this end, let us fix the parameters $(\tilde{G},\gamma_m,\gamma_0,\Gamma)$ and view $\omega_c$ (or equivalently, $\omega_d$, since $L_p$ and $\ell$ are fixed) as the unknown parameter to be sensed. For a given threshold value $\omega_c^*$, we can define the corresponding exceptional-point location $\mathbf{R}_{\rm EP}(\omega_c^*)$ by the expression (\ref{REP_omegaC}). Let us choose a closed loop $C$ in the parameter space, for instance, the circle
\begin{equation}
\bar{\Delta}(\theta) = \bar{\Delta}_c + R\cos\theta, \quad J(\theta) = J_c + R\sin\theta, \quad \theta\in[0,2\pi],
\label{loop_C_def}
\end{equation}
with center $(\bar{\Delta}_c,J_c)$ and radius $R$ chosen such that the point $\mathbf{R}_{\rm EP}(\omega_c^*)$ lies approximately on the loop, i.e., 
\begin{equation}
\big[\bar{\Delta}_0(\omega_c^*)-\bar{\Delta}_c\big]^2
+ \big[J_{\rm EP}(\omega_c^*)-J_c\big]^2 \simeq R^2.
\label{loop_threshold_condition}
\end{equation}
Thus, for $\omega_c$ slightly larger or smaller than $\omega_c^*$, the exceptional point moves to one side or the other of the loop $C$. By choosing the loop geometry appropriately, one can arrange that values $\omega_c>\omega_c^*$ correspond to the exceptional point lying inside $C$, while values $\omega_c<\omega_c^*$ correspond to the exceptional point lying outside $C$. The loop $C$ thus acts as a spatial (in the parameter space) comparator for the exceptional-point position and hence for $\omega_c$.

\vspace{2mm}

In practice, the loop (\ref{loop_C_def}) can be implemented by slowly modulating the control detuning $\bar{\Delta}$ and the intercavity coupling $J$ along the desired trajectory in the parameter space. The radius $R$ and center $(\bar{\Delta}_c,J_c)$ can be calibrated using independent measurements or numerical modeling of $\mathbf{R}_{\rm EP}(\omega_c)$. A single sensing cycle consists of the following steps:

\begin{enumerate}
\item \textbf{Initialization:} Setting the control parameters to a starting point on the loop, one applies a weak probe field to the passive cavity at varying detuning $\delta$ and records the transmission spectrum $T_b(\delta)\propto |J|^2/|D(\delta)|^2$ from the active cavity. Identifying the two resonance frequencies corresponding to the real parts of the optical-supermode eigenvalues, let us label them $\Omega_A(0)$ and $\Omega_B(0)$. At subsequent angles $\theta_k$, one identifies the branches $\Omega_A(\theta_k)$ and $\Omega_B(\theta_k)$ by continuity from their values at $\theta=0$, i.e., by following each resonance smoothly as a function of $\theta_k$, and not by reordering them by instantaneous frequency at each point.

\item \textbf{Encircling:} Let us now drive the control parameters $(\bar{\Delta}(t),J(t))$ slowly along the
loop $C$ defined by equation (\ref{loop_C_def}), with $\theta$ playing the role of a control phase. At a discrete set of angles $\theta_k\in[0,2\pi]$ ($k=1,\dots,N_\theta$), one measures the quasi-steady-state transmission spectrum $T_b(\delta)$ and extracts the two resonance branches $\Omega_{1,2}(\theta_k)$. By matching the peaks continuously as a function of $\theta_k$, one obtains two continuous branches, which one denotes $\Omega_A(\theta_k)$ and $\Omega_B(\theta_k)$, corresponding to the eigenvalues of $M_{\rm eff}(\bar{\Delta},J)$ along the loop. In practice, one ought to choose the loop at a buffer distance from $\mathbf{R}_{\rm EP}$ set by spectral resolution, so that the peaks remain distinguishable for all $\theta_k$. 

\item \textbf{Topological decision:} After one full loop ($\theta=2\pi$) the control parameters return to their
initial values. A measurement of $T_b(\delta)$ once more allows one to extract the final
resonance frequencies along the two branches, $\Omega_A(2\pi)$ and $\Omega_B(2\pi)$. Comparing $\Omega_{A,B}(0)$ with $\Omega_{A,B}(2\pi)$ can have two outcomes: (a) either $\Omega_A(2\pi)$ is the continuation of $\Omega_A(0)$ and $\Omega_B(2\pi)$ is the continuation of $\Omega_B(0)$ (no permutation of the branches), or (b) $\Omega_A(2\pi)$ is the continuation of $\Omega_B(0)$ and $\Omega_B(2\pi)$ is the continuation of $\Omega_A(0)$ (the branches are interchanged). If the branches remain unchanged, one can assign the digital outcome $\mathcal{Z} = 0$, indicating that the loop $C$ did not encircle the exceptional point, while if the branches are interchanged, one assigns $\mathcal{Z} = 1$, indicating that the loop $C$ did encircle the exceptional point.
\end{enumerate}

It may be emphasized that in this protocol, the loop $C$ is traversed in a quasi-static (slow) manner, i.e., at each angle $\theta_k$, the system is allowed to attain a steady state and the transmission spectrum $T_b(\delta)$ is then measured. This sequence of static measurements realizes an encircling of the exceptional point in parameter space without relying on nonadiabatic dynamical evolution along the loop. The observed permutation or nonpermutation of the eigenvalue branches is therefore a robust topological property of the stationary spectrum, rather than a dynamical consequence.

\vspace{2mm}

Since the loop $C$ was designed such that the exceptional point lies inside $C$ for $\omega_c>\omega_c^*$ and outside for $\omega_c<\omega_c^*$, $\mathcal{Z}=1$ implies $\omega_c>\omega_c^*$ while $\mathcal{Z}=0$ implies $\omega_c<\omega_c^*$. A single encircling experiment thus realizes a digital comparator for $\omega_c$ with threshold $\omega_c^*$. The same procedure works for sensing $\omega_d$ and, in fact, for a given $\ell$ determines whether $L_p$ is above or below a given threshold $L_p^*$. Because the outcome is a binary topological property, i.e., swap or no swap of the supermodes, the scheme is robust against small parameter perturbations that do not move the exceptional point across the loop boundary or compromise branch resolvability. This stands in contrast to continuous exceptional-point-based sensing, where the same mechanism that enhances the signal also enhances the impact of noise near the exceptional point \cite{Wiersig_2020,Lau_2018,Langbein_2018}.

\vspace{2mm}

A realistic implementation of the above-mentioned protocol requires two important considerations. First, the rate at which the control parameters $(\bar{\Delta}, J)$ are altered should be slow compared to the optical-relaxation timescale (although faster than the supercurrent lifetime), thereby ensuring that the transmission spectrum at each intermediate angle faithfully reflects the stationary eigenstructure of $M_{\rm eff}(\bar{\Delta},J)$. Second, for reliable branch tracking, the two resonances must remain spectroscopically resolvable except in a vanishingly-small neighborhood of the exceptional point so that peak identification is not compromised. Under these conditions, the discrete permutation or nonpermutation of the optical supermodes becomes a robust topological indicator of whether the exceptional point lies inside or outside the chosen loop.

\vspace{2mm}

Let us also remark that while a single encircling implements a binary topological test that determines whether the exceptional point lies inside or outside a prescribed contour in the control-parameter space, integer-level resolution of the winding number can be achieved by employing a sequence of such loops with systematically-shifted radii or centers. Each loop acts as a comparator with a distinct threshold value of $L_p$, determined by the corresponding exceptional-point location $\mathbf{R}_{\rm EP}(L_p)$. By combining the binary outcomes of multiple encircling measurements successively, one can distinguish between adjacent winding numbers $L_p$ and $L_p+1$, provided that the exceptional-point displacement associated with a unit change in $L_p$ exceeds the uncertainty in the loop boundary. This digital and topological approach therefore enables unit-resolution sensing while retaining robustness against small parameter fluctuations. By relying on a topological, nondestructive readout, this method offers a promising route for overcoming the intrinsic limitations of destructive matter-wave interferometry and enabling robust sensing of angular momentum even in the high-$L_p$ regime \cite{Eckel_2014,Pandey_2019}.

\vspace{2mm}

Finally, let us conclude this discussion by re-emphasizing that conventional exceptional-point-based sensing infers an unknown perturbation $\epsilon$ from the continuous eigenvalue splitting near an exceptional point, i.e., $\Delta\Omega(\epsilon)= {\rm Re}(\lambda_+-\lambda_-)\propto \sqrt{\epsilon}$, but the same mechanism also amplifies fluctuations, thereby limiting practical advantage in the presence of technical and quantum noise \cite{Wiersig_2020,Lau_2018,Langbein_2018}. In contrast, our digital protocol does not estimate $\epsilon$ from resolving a small splitting and instead relies on the topological permutation or nonpermutation of the eigenbranches under a closed loop. The outcome depends only on whether the exceptional point lies inside or outside the loop and is therefore robust to small parameter noise that deforms the loop without changing its winding relative to the exceptional point. A limiting factor of our protocol is therefore the possibility of a topological misclassification that can occur only if fluctuations displace the exceptional point across the loop boundary, which may, however, be avoided by a careful calibration of the loop.

\section{Conclusions}\label{conc_sec}
In this work, we have demonstrated that atomic backaction from a ring-trapped BEC provides a natural modification of non-Hermitian dimer physics. By deriving an exact Schur-complement reduction of the full light-matter dynamics, we identified how the atomic sidemodes induce a complex self-energy that renormalizes the optical detuning of the passive cavity and the gain-loss balance, producing a tunable exceptional point. The measurable consequences of this renormalization appear directly in the transmission spectrum where the modified eigenvalues govern the resonance structure. Building on the associated half-integer topological charge, we introduced a digital-sensing protocol based on the permutation of the optical supermodes under encircling of the exceptional point, thereby providing a topological binary-readout alternative to continuous exceptional-point-based sensing, with robustness against small perturbations that do not alter the loop’s winding relative to the exceptional point. Our results demonstrate that cavity-BEC platforms may serve as reconfigurable non-Hermitian photonic systems and offer a unified route to exceptional-point control, spectroscopy, and topological sensing within a single architecture. \\

\textbf{Acknowledgements:} A.G. gratefully acknowledges discussions with Bijan Bagchi, Miloslav Znojil, Akash Sinha, and Avinash Khare on $\mathcal{PT}$-symmetric systems. M.B. thanks the Air Force Office of Scientific Research (AFOSR) (FA9550-23-1-0259) for support.

\begin{widetext}

\appendix 
\section{Schur-complement reduction}\label{appA}
The matrix (\ref{Lambdadef}) admits the block form
\begin{equation}
\Lambda =
\begin{pmatrix}
\mathcal{A} & \mathcal{B} \\[4pt]
\mathcal{C} & \mathcal{D}
\end{pmatrix},
\end{equation} where 
\begin{equation}
\mathcal{A}=
\begin{pmatrix}
\bar{\Delta}+i\dfrac{\gamma_0}{2} & -J \\[3pt]
-J & \bar{\Delta}-i\dfrac{\Gamma}{2}
\end{pmatrix},
\end{equation} is the optical block acting on the subspace $(a,b)^T$, 
\begin{equation}
\mathcal{D}=
\begin{pmatrix}
-\omega_c+i\dfrac{\gamma_m}{2} & 0 \\[3pt]
0 & -\omega_d+i\dfrac{\gamma_m}{2}
\end{pmatrix},
\label{block_decomposition}
\end{equation} is the atomic block acting on the subspace $(c,d)^T$, and the light-matter coupling is given by the blocks
\begin{equation}
\mathcal{B}=
\begin{pmatrix}
-\tilde{G} & -\tilde{G}\\[3pt]
0 & 0
\end{pmatrix},\quad \quad
\mathcal{C}=\mathcal{B}^T.
\end{equation}
The right-eigenvalue equation $\Lambda v = \lambda v$ separates as
\begin{eqnarray}
(\mathcal{A}-\lambda I)v_A + \mathcal{B} v_D &=& 0, \label{eig1}\\
\mathcal{C} v_A + (\mathcal{D}-\lambda I)v_D &=& 0, \label{eig2}
\end{eqnarray}
where $v_A$ and $v_D$ represent, respectively, the optical and atomic components of the eigenvector. Equation (\ref{eig2}) implies
\begin{equation}
v_D = - (\mathcal{D}-\lambda I)^{-1} \mathcal{C} v_A.
\end{equation}
Substitution into equation (\ref{eig1}) leads to the following effective $2\times2$ non-Hermitian eigenproblem for $v_A$:
\begin{equation}
\big[\mathcal{A} - \lambda I - \mathcal{B} (\mathcal{D}-\lambda I)^{-1} \mathcal{C}\big] v_A = 0,
\end{equation}
defining the following exact Schur-complement reduction:
\begin{equation}
M_{\rm eff}(\lambda) = \mathcal{A} - \mathcal{B} (\mathcal{D}-\lambda I)^{-1} \mathcal{C}.
\label{Meff_def}
\end{equation}
Since $\mathcal{D}$ is diagonal, we can write
\begin{equation}
(\mathcal{D}-\lambda I)^{-1} = {\rm diag}\left(\dfrac{1}{-\omega_c+i\dfrac{\gamma_m}{2}-\lambda},\dfrac{1}{-\omega_d+i\dfrac{\gamma_m}{2}-\lambda}\right),
\end{equation}
and therefore the resulting effective $2 \times 2$ matrix acting on the optical subspace assumes the expression quoted in equation (\ref{Mefflambda}) with complex self-energy (\ref{selfenergy}). 

\section{Static approximation}\label{appB}
In the exact Schur-complement reduction, the atomic backaction on the optical subspace enters through the frequency-dependent self-energy (\ref{selfenergy}), where $\lambda$ denotes a complex eigenvalue of the full non-Hermitian matrix (\ref{Lambdadef}). For analytical tractability, we worked in a regime where the optical eigenvalues are close to a chosen control detuning $\bar{\Delta}$, while remaining far away from
the atomic-sidemode resonances at $-\omega_{c}$ and $-\omega_{d}$.  In this case, it is natural to approximate $\Sigma(\lambda)$ by its static value $\Sigma(\bar{\Delta})$ evaluated at the optical detuning. A convenient way to make this approximation precise is to expand $\Sigma(\lambda)$ about $\lambda=\bar{\Delta}$ in the manner
\begin{equation}
\Sigma(\lambda) = \Sigma(\bar{\Delta}) + \Sigma'(\bar{\Delta})(\lambda-\bar{\Delta}) + \mathcal{O}\big((\lambda-\bar{\Delta})^{2}\big),
\end{equation}
with
\begin{equation}
\Sigma'(\bar{\Delta}) = -\frac{\tilde{G}^{2}}{\big(\bar{\Delta}+\omega_{c}-i\gamma_{m}/2\big)^{2}}
-\frac{\tilde{G}^{2}}{\big(\bar{\Delta}+\omega_{d}-i\gamma_{m}/2\big)^{2}}.
\end{equation}
The static approximation $\Sigma(\lambda)\simeq\Sigma(\bar{\Delta})$ is valid provided the linear correction is small compared to the leading term, i.e., 
\begin{equation}
\big|\Sigma'(\bar{\Delta})(\lambda-\bar{\Delta})\big| \ll \big|\Sigma(\bar{\Delta})\big|.
\label{static_error_criterion}
\end{equation}
Near $\bar{\Delta}_0=-(\omega_c+\omega_d)/2$, where the dispersive part of the self-energy vanishes while the imaginary part remains finite, this condition should be understood termwise for each atomic susceptibility. Using the explicit forms above, this condition can be expressed in a transparent way. Each contribution to the self-energy has the structure $\tilde{G}^{2}/(\bar{\Delta}+\omega_{j}-i\gamma_{m}/2)$ with
$j=c,d$, so that the relevant small parameter is
\begin{equation}
\epsilon_{j} = \frac{|\lambda-\bar{\Delta}|}{|\bar{\Delta}+\omega_{j}-i\gamma_{m}/2|}, \quad \quad j=c,d.
\end{equation}
If
\begin{equation}
|\bar{\Delta}+\omega_{j}-i\gamma_{m}/2| \simeq \sqrt{(\bar{\Delta}+\omega_{j})^{2}+(\gamma_{m}/2)^{2}}
\gg |\lambda-\bar{\Delta}|,
\label{static_condition}
\end{equation}
then $\epsilon_{j}\ll 1$ and the relative error in equation (\ref{static_error_criterion}) is parametrically small.  Since
$\gamma_{m}$ is much smaller than all optical scales, the condition (\ref{static_condition}) is effectively controlled by the real detuning $|\bar{\Delta}+\omega_{j}|$.  In particular, for the parameter regime of interest, we have chosen $\bar{\Delta}$ such that $|\bar{\Delta}+\omega_{c,d}|$ remains of the order of a few $\gamma_{0}$ or larger, while the optical-eigenvalue splitting is of order $J\lesssim\gamma_{0}$. This ensures that $|\bar{\Delta}+\omega_{c,d}|\gg|\lambda-\bar{\Delta}|$
and $|\bar{\Delta}+\omega_{c,d}|\gg\gamma_{m}/2$, so that the
frequency-dependence of $\Sigma(\lambda)$ over the relevant optical bandwidth is negligible. Under these conditions, it is justified to replace $\Sigma(\lambda)$ by its static value $\Sigma(\bar{\Delta})$ and to work with the effective $2\times 2$ matrix $M_{\rm eff}(\bar{\Delta})$ in the analysis of the optical eigenvalues and the transmission spectrum.

\vspace{2mm}

A potential concern may be that whether the $\lambda$-dependence of $\Sigma(\lambda)$ could qualitatively modify, smear, or remove the exceptional point. In the present setting, the relevant control parameter is the spectral distance between the optical window and the atomic poles at $\lambda\simeq -\omega_{c,d}+i\gamma_m/2$. Provided that the exceptional point is engineered in an off-resonant region where these poles are not approached, the self-energy remains analytic and slowly varying, and its frequency dependence may produce only perturbative shifts to the exceptional-point location. More precisely, if $\Sigma(\lambda)$ is weakly dispersive over the optical window, one may treat the difference $\delta\Sigma(\lambda) = \Sigma(\lambda)-\Sigma(\bar{\Delta})$ as a small perturbation. Thus the magnitude of this perturbation is bounded by
\begin{equation}
|\delta\Sigma(\lambda)| \le |\Sigma'(\bar{\Delta})||\lambda-\bar{\Delta}|
+\mathcal{O}(|\lambda-\bar{\Delta}|^2).
\end{equation}
Under the same small-parameter condition $\epsilon_j\ll 1$ in equation (\ref{static_condition}), the induced changes in the coefficients of the characteristic polynomial are perturbatively small. Consequently, within the two-dimensional control manifold $(\bar{\Delta},J)$ and in the off-resonant regime $\epsilon_j\ll 1$, the exceptional-point condition (discriminant $=0$) is perturbed smoothly, i.e., the vanishing of the discriminant is preserved under the small analytic correction $\delta\Sigma(\lambda)$, thereby implying that the exceptional point persists but its location in $(\bar{\Delta},J)$ is shifted by a small amount of the order of $|\delta\Sigma|$.

\section{Input noise and susceptibility denominator in the static regime}\label{appC}

In the static regime, the reduced optical subsystem obeys linear quantum Langevin equations
\begin{equation}
\dot{A}_{\rm opt}= i M_{\rm eff}A_{\rm opt} + A_{\rm opt,in}(t), 
\label{QLE_optical}
\end{equation}
where $A_{\rm opt}=(a,~b)^T$, $M_{\rm eff}=M_{\rm eff}(\bar{\Delta})$ is given by equation (\ref{Meff_static}), and $A_{\rm opt,in}(t)$ collects the input-noise operators associated with the passive and active channels. In the frequency space, one can write (with possible constant phases absorbed)
\begin{equation}
A_{\rm opt}(\delta)= \left(M_{\rm eff}-\delta I\right)^{-1}A_{\rm opt,in}(\delta)+
\left(M_{\rm eff}-\delta I\right)^{-1}F_{\rm probe}(\delta),
\end{equation}
where $F_{\rm probe}=(\eta,~0)^T$ represents the probe drive. Thus the mean coherent response is governed by the same susceptibility denominator $D(\delta)=\det(M_{\rm eff}-\delta I)$ appearing in equation (\ref{probe_sol_expt}) with noise entering as additive fluctuations filtered by the same linear response. Most importantly, the topological-sensing protocol of Sec. (\ref{topology_sec}) relies on the eigenbranch permutation under a closed loop (a binary topological observable) and thus the dominant limitation is set by whether fluctuations change the inside or outside classification of the exceptional point relative to the chosen loop.

\section{Physical conditions for an exceptional point}\label{appD}
The condition (\ref{EP_condition}) can be expressed as
\begin{eqnarray}
J_{\rm EP}^2 &=& \frac{1}{16} \left[(\gamma_0 + \Gamma) - 2i({\rm Re}[\Sigma(\bar{\Delta})] + i {\rm Im}[\Sigma(\bar{\Delta})])\right]^2 \nonumber \\
&=& \frac{1}{16} \left[(\gamma_0 + \Gamma) - 2i{\rm Re}[\Sigma(\bar{\Delta})] +2 {\rm Im}[\Sigma(\bar{\Delta})]\right]^2 \nonumber \\
&=& \frac{1}{16} \left[(\gamma_0 + \Gamma +2 {\rm Im}[\Sigma(\bar{\Delta})]) - 2i{\rm Re}[\Sigma(\bar{\Delta})]\right]^2 \nonumber \\
&=& \frac{1}{16} \left[(\gamma_0 + \Gamma +2 {\rm Im}[\Sigma(\bar{\Delta})])^2 -4{\rm Re}[\Sigma(\bar{\Delta})]^2 - 4i {\rm Re}[\Sigma(\bar{\Delta})](\gamma_0 + \Gamma +2 {\rm Im}[\Sigma(\bar{\Delta})]) \right]. 
\label{EPgencondition1}
\end{eqnarray}
Since the left side is real, we have our first condition 
\begin{equation}
{\rm Re}[\Sigma(\bar{\Delta})](\gamma_0 + \Gamma +2 {\rm Im}[\Sigma(\bar{\Delta})]) = 0.
\end{equation}

\subsection*{Case 1}
Taking ${\rm Re}[\Sigma(\bar{\Delta})] \neq 0$, one finds that this condition can be met only when
\begin{equation}
\gamma_0 + \Gamma = -2 {\rm Im}[\Sigma(\bar{\Delta})].
\label{realitycase1}
\end{equation}
If this condition is met, an exceptional point occurs at
\begin{equation}
J_{\rm EP} = \frac{1}{4} \sqrt{(\gamma_0 + \Gamma +2 {\rm Im}[\Sigma(\bar{\Delta})])^2 -4{\rm Re}[\Sigma(\bar{\Delta})]^2},
\end{equation} and combining with the reality condition (\ref{realitycase1}), one gets
\begin{equation}
J_{\rm EP} = \frac{1}{4} \sqrt{ -4{\rm Re}[\Sigma(\bar{\Delta})]^2},
\end{equation} i.e., it is only satisfied in the trivial case $J_{\rm EP} = 0$ due to the reality of $J_{\rm EP}$. So one does not get any nontrivial exceptional point in this case. 

\subsection*{Case 2}
For nontrivial exceptional points, we must first have
\begin{equation}
{\rm Re}[\Sigma(\bar{\Delta}_0)] = 0,
\end{equation}
which means from expression (\ref{ReSigma}), that the following condition must be met:
\begin{equation}
\bigg[\frac{\bar{\Delta}_0 + \omega_c}{(\bar{\Delta}_0 + \omega_c)^2 + (\gamma_m/2)^2} + \frac{\bar{\Delta}_0 + \omega_d}{(\bar{\Delta}_0 + \omega_d)^2 + (\gamma_m/2)^2} \bigg] = 0, 
\end{equation} giving us
\begin{equation}
(\bar{\Delta}_0 + \omega_c)[(\bar{\Delta}_0 + \omega_d)^2 + (\gamma_m/2)^2] + (\bar{\Delta}_0 + \omega_d)[(\bar{\Delta}_0 + \omega_c)^2 + (\gamma_m/2)^2] =0.
\end{equation}
This can be factorized as 
\begin{equation}
(2\bar{\Delta}_0 + \omega_c + \omega_d)[(\bar{\Delta}_0 + \omega_c)(\bar{\Delta}_0 + \omega_d) + (\gamma_m/2)^2]=0.
\end{equation}
So we will have three possible values of $\bar{\Delta}=\bar{\Delta}_0$ satisfying this. From the first factor, we have
\begin{equation}
\bar{\Delta}_0 = -\frac{(\omega_c + \omega_d)}{2},
\label{middleroot}
\end{equation} while from the second factor, we have
\begin{equation}
\bar{\Delta}_0 = -\frac{(\omega_c + \omega_d)}{2} \pm \frac{1}{2} \sqrt{(\omega_c - \omega_d)^2 - \gamma_m^2} \simeq -\omega_c,-\omega_d,
\label{pmroot}
\end{equation} since $|\omega_c - \omega_d| \gg \gamma_m$. However, since $\bar{\Delta} \simeq - \omega_{c,d}$ is near the atomic resonances for which the static approximation is compromised, we will focus on the middle root (\ref{middleroot}). If one substitutes the expression (\ref{middleroot}) into equation (\ref{ImSigma}), one gets
\begin{eqnarray}
{\rm Im}[\Sigma(\bar{\Delta}_0)] &=& \tilde{G}^2 \bigg[ \frac{4 \gamma_m}{(\omega_c - \omega_d)^2 + \gamma_m^2} \bigg] \simeq  \frac{4 \tilde{G}^2 \gamma_m}{(\omega_c - \omega_d)^2}. 
\label{ImSigma_middleroot}
\end{eqnarray}
Then the exceptional point is simply given by the condition (\ref{EPgencondition1}) as
\begin{eqnarray}
J_{\rm EP} = \frac{1}{4} (\gamma_0 + \Gamma + 2 {\rm Im}[\Sigma(\bar{\Delta}_0)]),
\end{eqnarray}
which using the expression (\ref{ImSigma_middleroot}) agrees with the condition (\ref{JEP_condition_physical}) quoted in the main text. 

\section{Topological charge of the exceptional point}\label{appE}
To derive the standard result that the exceptional point has a topological charge of $1/2$ up to a sign that depends on the orientation of the loop, let us expand near the exceptional point as
\begin{equation}
\mathbf{R} = \mathbf{R}_{\rm EP} + \delta\mathbf{R}, \quad \quad \delta\mathbf{R} = (\delta\bar{\Delta},\delta J),
\end{equation}
and linearize the discriminant as
\begin{equation}
\mathfrak{D}(\mathbf{R}) \simeq \mathfrak{D}(\mathbf{R}_{\rm EP}) + \nabla_{\mathbf{R}}\mathfrak{D}(\mathbf{R}_{\rm EP})\cdot\delta\mathbf{R} = \mathbf{u}\cdot\delta\mathbf{R},
\quad \quad \mathbf{u} = \nabla_{\mathbf{R}}\mathfrak{D}(\mathbf{R}_{\rm EP}),
\label{D_linearized}
\end{equation}
where we have used $\mathfrak{D}(\mathbf{R}_{\rm EP})=0$. The complex vector $\mathbf{u}$ essentially encodes the local sensitivity of the discriminant to deviations in $\bar{\Delta}$ and $J$. The eigenvalue splitting near the exceptional point reads
\begin{equation}
\lambda_+(\mathbf{R})-\lambda_-(\mathbf{R}) \simeq \frac{1}{2}
\sqrt{\mathfrak{D}(\mathbf{R})} \simeq \frac{1}{2} \sqrt{\mathbf{u}\cdot\delta\mathbf{R}},
\label{EP_splitting_local}
\end{equation}
with the branch cut chosen consistently. Let us now consider a small closed loop $C$ in the parameter space that encircles the exceptional point once. A convenient parametrization is a circle of radius $R$ as given by
\begin{equation}
\bar{\Delta}(\theta) = \bar{\Delta}_0 + R\cos\theta, \quad \quad J(\theta) = J_{\rm EP} + R\sin\theta, \quad \quad \theta\in[0,2\pi].
\label{loop_def}
\end{equation}
Substitution of the parametric expressions (\ref{loop_def}) into the expansion (\ref{D_linearized}) yields
\begin{equation}
\mathfrak{D}(\theta) \simeq \mathbf{u}\cdot \big(R\cos\theta, R\sin\theta\big) = R|\mathbf{u}\cdot(\cos\theta,\sin\theta)|e^{i\varphi(\theta)},
\end{equation}
where the phase $\varphi(\theta)=\arg[\mathfrak{D}(\theta)]$ winds by $2\pi$ as $\theta$ goes from $0$ to $2\pi$, provided the exceptional point lies inside the loop, which is assumed to be small. The eigenvalue difference then acquires the characteristic square-root dependence
\begin{equation}
\lambda_+(\theta)-\lambda_-(\theta) \simeq \frac{1}{2}\sqrt{\mathfrak{D}(\theta)} = \frac{1}{2}\sqrt{R|\mathbf{u}\cdot(\cos\theta,\sin\theta)|}
\exp\big[i\varphi(\theta)/2\big].
\end{equation}
As a consequence, under a single loop around the exceptional point, we have
\begin{equation}
\lambda_+(\theta=2\pi)-\lambda_-(\theta=2\pi) = -\big[\lambda_+(\theta=0)-\lambda_-(\theta=0)\big],
\end{equation}
so that the individual eigenvalues are permuted, i.e., 
\begin{equation}
\lambda_+(\theta=2\pi) = \lambda_-(\theta=0), \quad \quad \lambda_-(\theta=2\pi) = \lambda_+(\theta=0).
\label{EP_permutation}
\end{equation}
A convenient measure of the topological charge is the winding of the phase of the complex energy difference \cite{Berry_2004,Heiss_2012,Wiersig_2020}, given by
\begin{equation}
q_{\rm EP} = \frac{1}{2\pi} \oint_C \nabla_{\mathbf{R}} \arg\big[\lambda_+(\mathbf{R}) - \lambda_-(\mathbf{R})\big] \cdot d\mathbf{l}.
\label{qEP_def}
\end{equation}
Using $\lambda_+-\lambda_-=\tfrac{1}{2}\sqrt{\mathfrak{D}}$ and $\arg[\sqrt{z}]=\tfrac{1}{2}\arg[z]$, this reduces to
\begin{equation}
q_{\rm EP} = \frac{1}{4\pi} \oint_C \nabla_{\mathbf{R}}
\arg\big[\mathfrak{D}(\mathbf{R})\big] \cdot d\mathbf{l} = \frac{\Delta\arg[\mathfrak{D}]}{4\pi},
\end{equation}
with $\Delta\arg[\mathfrak{D}]$ being the total change in the argument of the discriminant along the loop. For a small loop encircling the exceptional point once, $\Delta\arg[\mathfrak{D}]=2\pi$ and therefore $q_{\rm EP}$ is given by the expression (\ref{halfint_charge}) quoted in the main text. This half-integer charge is directly manifested in the eigenvalue permutation (\ref{EP_permutation}), forming the basis of our digital-sensing protocol.

\end{widetext}

\end{document}